\documentclass[%
 reprint,
 amsmath,amssymb,
 aps,
prb,
]{revtex4-1}

\usepackage{graphicx}% Include figure files
\usepackage{dcolumn}% Align table columns on decimal point
\usepackage{bm}% bold math
\begin{document}

\preprint{APS/123-QED}

\title{Mechanisms of nonlinear spin-wave emission from a microwave driven nanocontact}% Force line breaks with \\

\author{Florin Ciubotaru}
  \email{ciubotaru@physik.uni-kl.de}
\author{Alexander A. Serga}%
\author{Britta Leven}
\author{Burkard Hillebrands}
\affiliation{%
 Fachbereich Physik and Landesforschungszentrum OPTIMAS, Technische Universit\"{a}t Kaiserslautern, 67663 Kaiserslautern,
 Germany}%

\author{Luis Lopez Diaz}
\affiliation{
 Departamento de Fisica Aplicada, University of Salamanca, 37008 Salamanca, Spain}%

\date{\today}% It is always \today, today,

\begin{abstract}
We present a micromagnetic study of linear and nonlinear spin-wave
modes excited in an extended permalloy thin film by a microwave
driven nanocontact. We show that the linear mode having the
frequency equal to the excitation frequency (\emph{f}) is driven
by the ac Oersted field component perpendicular to the static
external field (applied in-plane of the sample). The nonlinear
mode with the frequency \emph{f}/2 is excited as an independent
eigenmode within a parametric longitudinal pumping process (due ac
Oersted field component parallel to the bias field). Spectral
positions of those modes are determined both in the space and
phase domain. The results are important for the transfer of
information coded into spin-waves between nanocontacts, and for
synchronization of spin transfer torque nano-oscillators.

%The \emph{f} mode is generated at the edge of the contact and is
%propagating nearly perpendicular to the bias field. The \emph{f}/2
%mode is localized outside the contact area and is propagating
%under a small angle with respect to the bias field.

\begin{description}
\item[PACS numbers]{75.30.Ds, 75.78.Cd, 75.75.+a}
\end{description}
\end{abstract}

\maketitle

Spin waves in meso- and micro-sized magnetic structures have
attracted growing attention due to the prospective applications in
magnon spintronic devices. For example, the use of phase and
amplitude of spin waves as an information carrier in magnetic
logic circuits is extensively
studied.\cite{Kostylev,Kim,Schneider,Khitun} The problem of
spin-wave excitation is of crucial importance for the
miniaturization of such devices to the nano-scale. The advanced
research on device miniaturization makes it possible to study the
spin waves exited by microwave driven nanosized antennas
(nanocontacts). \cite{Tsoi,Demidov,Schultheiss,Balkashin} This
task is not trivial: if the density of the microwave current is
high enough, the spin-wave system is pushed far beyond the linear
regime, and nonlinear processes appear. Moreover, in nano-sized
structures, the properties like the threshold power and frequency
limit of the nonlinear excited spin-wave modes can be tuned by a
superimposed direct current flowing through the
contact.\cite{Schultheiss} Vice versa, the influence of the
microwave currents on the magnetic oscillations caused by the spin
transfer torque effect (STT) in direct current driven nanocontacts
was demonstrated.\cite{Rippard,Florez} It was shown that the
propagation length of the STT emitted spin waves can be enhanced
by using an additional microwave current injected through the
contact.\cite{Demokritov} The microwave current plays also an
important role in the synchronization of STT based
nano-oscillators (STNO).\cite{Rippard,Georges,Urazhdin} All these
tasks need a deeper understanding of the characteristics of spin
waves emitted by microwave driven nanocontacts.

In this work we performed systematic micromagnetic simulations
\cite{OOMMF} on magnetization dynamics in a nanocontact structure
to reveal the underlaying mechanism of magnons excitation and to
characterize the linear and nonlinear spin-wave modes detected
experimentally in such devices.\cite{Schultheiss}

\begin{figure}[b]
\includegraphics[width=8.5 cm,clip]{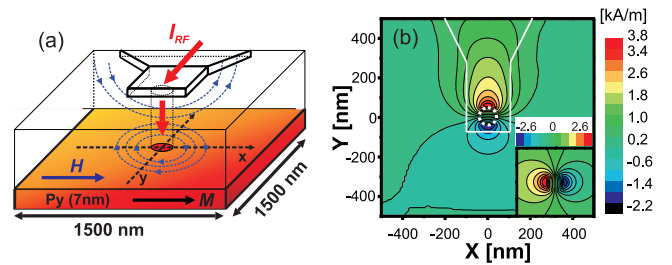}% Here is how to import EPS art
\caption{\label{fig:structure} (a) Scheme of the sample with the
contact in the center and a nonmagnetic metallic asymmetric lead
situated at 60 nm above the magnetic layer. (b) Spatial
distribution of the Oersted field components parallel and
perpendicular (inset) to the external applied field. The inset
figure has a lateral size of 300 nm $\times$ 300 nm.}
\end{figure}

In the simulations we considered a thin permalloy layer of 7 nm
thickness with lateral dimensions of $1.5\times1.5$~$\mu$m$^{2}$
(see Fig.~\ref{fig:structure}(a)). As in the experiment
\cite{Schultheiss}, an in-plane magnetic field of 19.89~kA/m (250
Oe) was applied along the $x$-axis. A microwave current is
injected through the nanocontact placed in the center of the
structure as defined in Fig.~\ref{fig:structure}(a). The resulting
alternating Oersted field excites magnetization precession in the
system. The circular shape contact has a diameter of 100~nm. The
standard material parameters of permalloy used to simulate the
magnetization dynamics are: saturation magnetization
$\mu_{0}M_{S}=1$~T, exchange stiffness constant $A =
1.3\times10^{-11}$~J/m and zero magnetocystalline anisotropy was
assumed. The simulated area was discretized into $N_{x}\times
N_{y}\times N_{z}=300\times300\times2$ cells, each cell having a
size of $5\times5\times3.5$~nm$^{3}$. Initially, the magnetization
is aligned in-plane along the $x$-axis. To model the parametric
processes in our system, we considered a random fluctuation of the
magnetization within 0.5$^{\circ}$ with respect to the $x$-axis,
which is caused by thermal noise in real systems. The damping
boundary conditions were used to avoid spin-wave reflection at the
edges of the simulation area. A spatial distribution of the
damping parameter ($\alpha(x,y)$) in the Landau-Lifshitz-Gilbert
equation (with the STT term included) given by: $\alpha =
\alpha_{0} + \lambda \Big[1 + \textnormal{arctan} \Big(\frac
{\sqrt{x^{2}+y^{2}} - R} {\sigma} \Big)\Big]$ with $\lambda = 1$,
$R = 800$ nm, and $\sigma = 100$~nm ensures a constant value
$\alpha_{0} = 0.01$ in the center of the sample for a radius of
$\sim 500$ nm. The variables $x$ and $y$ denote the spatial
coordinates. With the above parameters, the damping is increased
more than fifty times at the edges of the simulation area, thus
the spin-wave intensity decays strongly towards the boundaries. We
calculate \cite{LLG} the current distribution and the
corresponding Oersted field in the Permalloy layer with and
without the contribution of an asymmetric top electrode (according
to the real structure in the experiment \cite{Schultheiss}). It
was found that the lead induces an asymmetry for the Oersted field
component parallel to the external field while the perpendicular
component remains virtually unchanged
(Fig.~\ref{fig:structure}(b)). In the presented simulations we
used the distribution of the Oersted field with the contribution
of the lead included.

From the experiment \cite{Schultheiss} it was found that two
conditions must be fulfilled in order to generate a spin-wave mode
with half of the excitation frequency. Firstly, the excitation
frequency must be at least twice the frequency at the bottom of
the spin-wave dispersion band (noted $\emph{f}_{0}$), and
secondly, the applied microwave power is above a certain threshold
value. In our configuration $\emph{f}_{0} \sim 4.55$ GHz, we set
the excitation frequency to 11 GHz to satisfy the first condition.
The exact value for the threshold power could not be determined by
simulations due to extremely long computation times. We used an
applied microwave power sufficiently large in order to observe the
nonlinear excited spin waves with good contrast.

\begin{figure}[b]
\includegraphics[width=8.5 cm,clip]{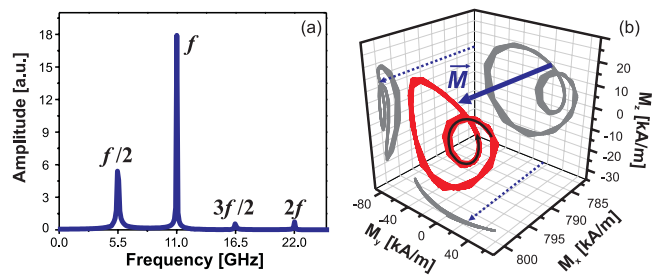}
\caption{\label{fig:spectrum} (a) Magnetization oscillation
spectrum for a cell at the top edge of the contact at the
excitation frequency \emph{f}~=~11~GHz. (b) Trajectory of the
local magnetization vector tip for the same cell at the edge of
the contact.}
\end{figure}

Analyzing the magnetization fluctuation in time for each
computational cell and performing the Fast Fourier Transform
(FFT), we obtained the frequency spectra for each component of the
magnetization. The behavior of the magnetization in one
computational cell (top edge of the contact) is presented in
Fig.~\ref{fig:spectrum}. One sees a response with the same
frequency as the excitation one (\emph{f}), but also the
generation of nonlinear modes: 0.5\emph{f}, 1.5\emph{f} and
2\emph{f} (Fig.~\ref{fig:spectrum}(a)). Furthermore, the
precession of the magnetization vector tip has a steady
complicated trajectory when the nonlinear modes are generated (see
Fig.~\ref{fig:spectrum}(b)), different from typical clamshell
orbits found in thin films for linear excitations. An additional
small loop was observed on the main gyration motion. This
peculiarity originates from the combination of partial
magnetization motions at different frequencies.

Our main task is to identify the excitation mechanism, spatial
localization and the wavevector directions as well as wave numbers
for both linear and nonlinear excited spin-waves modes.

\begin{figure}[b]
\includegraphics[width=8.5 cm,clip]{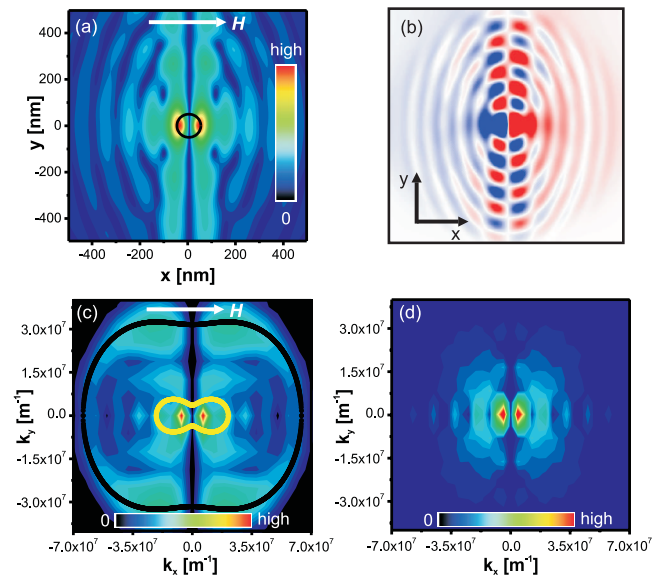}
\caption{\label{fig:fmode} (a) Spatial distribution of the
\emph{f} mode and the position of the contact in the center (black
circle). (b) Snapshot image of the magnetization oscillation
pattern with the component $M_{z}$ color coded (red - positive
values, blue - negative values). (c) Distribution of the \emph{f}
mode in the wavevector space. The black and yellow lines represent
the constant-frequency curves calculated analytical for 11 GHz and
5.5 GHz, respectively. (d) Distribution of the perpendicular
component of the Oersted field computed in the wavevector space.}
\end{figure}

First, we address the generation of the linear mode for which the
frequency matches the microwave one (\emph{f}). The spatial
distribution is obtained by taking the amplitude value at \emph{f}
= 11 GHz from the frequency spectra of each computational cell and
displaying them as a function of the position. The obtained map is
shown in Fig.~\ref{fig:fmode}(a). The black circle in the center
represents the position of the nanocontact. One can observe two
maxima that are located on the left and right side edges of the
contact. This can be understood by taking into account that the
main driving field at the excitation frequency is the Oersted
field component perpendicular to the bias field, with a very
similar amplitude distribution in space (see the inset from
Fig.~\ref{fig:structure}(a)). From Fig.~\ref{fig:fmode}(a), we
identified two types of waves excited by the nanocontact:
travelling and non-propagating waves. The travelling spin waves
exhibit in four beams starting from those two maxima regions, and
are propagating along directions nearly perpendicular to the bias
field (applied on the $x$-axis), while the non-propagating ones
can be observed in the rest of space. Moreover, the two maxima
regions are acting as two independent sources which emit spin
waves out of phase (180$^{\circ}$) in respect to each other. This
is illustrated on the snapshot image of the magnetization
oscillation pattern in the entire permalloy layer
(Fig.~\ref{fig:fmode}(b)) with the $M_{z}$ component color coded.
Since the two spin-wave beams that are travelling on the same side
of the contact (up or down) are out of phase, a destructive
interference takes place between them. Consequently, the spin-wave
intensity is strongly reduced on the direction perpendicular to
the bias field in the center of the contact
(Figs.~\ref{fig:fmode}(a) and ~\ref{fig:fmode}(b)).

The wavevectors of the emitted spin waves are determined by
applying the two-dimensional Fourier transform in space to the
already calculated components (real and imaginary parts) of the
temporal FFT for each computational cell. Figure
\ref{fig:fmode}(c) shows the distribution of the \emph{f} mode in
the wavevector space. $k_{x}$ and $k_{y}$ axis are the wave
numbers that indicate the in-plane directions of propagation
parallel and perpendicular to the external applied magnetic field,
respectively. The black and yellow contours in the figure
represent the analytical constant-frequency curve calculated from
the dispersion relation \cite{Kalinikos} for a frequency of 11 GHz
and 5.5 GHz, respectively. The wavevector data obtained from the
simulation is in very good agreement with the analytical
constant-frequency curve calculated at 11 GHz (the black line in
the figure).

The analysis in the wavevector space (Fig.~\ref{fig:fmode}(c))
shows that indeed the most efficiently excited spin waves are the
ones that are propagating with wavevectors nearly perpendicular
with respect to the bias field. The excitation efficiency is
decreasing fast with the spin-wave propagation angle, becoming
close to zero for spin waves with wavevectors parallel to the bias
field. This is associated with the known effect of decreasing
excitation efficiency for spin-waves with short wavelengths.
Nevertheless, the extracted data from the simulation shows another
two strong maxima located at small wave numbers. Those maxima are
related to the forced non-resonant driving of magnetization by the
excitation field. Their location is traced back to the spatial
configuration of the driving ac field. Figure \ref{fig:fmode}(d)
presents the distribution of the perpendicular component of the
Oersted field computed in the wavevector space. By comparing these
distributions (magnetization and excitation field) one observes
that the maxima are located at the same wave numbers. Thus, one
can see that the perpendicular component of the driving Oersted
field is responsible for excitation of both propagating and
localized spin-wave modes at the signal frequency \emph{f}. The
excitation efficiency for travelling waves reaches the maximum at
directions perpendicular to the bias field.

Next we discuss the process of excitation and the characterization
of the non-linear spin-wave modes, especially the \emph{f}/2 mode,
is described in the second part of the article.

The appearance of the \emph{f}/2 mode (= 5.5 GHz), can be
potentially attributed to three different excitation mechanisms.
Mechanism 1 is related to the three-magnon scattering process
(confluence and splitting of magnons - quanta of the spin waves
which are eigenexcitation of the magnetic system). Within
mechanism 2, the $H_{y}$ component of the driving Oersted field
excites an ac magnetization perpendicular to the bias field. This
ac magnetization serves as a pump for the \emph{f}/2 mode, via the
so-called process of parametric excitation of spin waves under
perpendicular pumping. Within this process, the \emph{f}/2 mode
can be generated directly from the forced excitation of
magnetization at the frequency \emph{f}. Mechanism 3 is a direct
excitation of the \emph{f}/2 mode by the Oersted field component
parallel to the bias field, a process called longitudinal (or
parallel) parametric pumping.\cite{Melkov}

\begin{figure}[t]
\includegraphics[width=8.5 cm,clip]{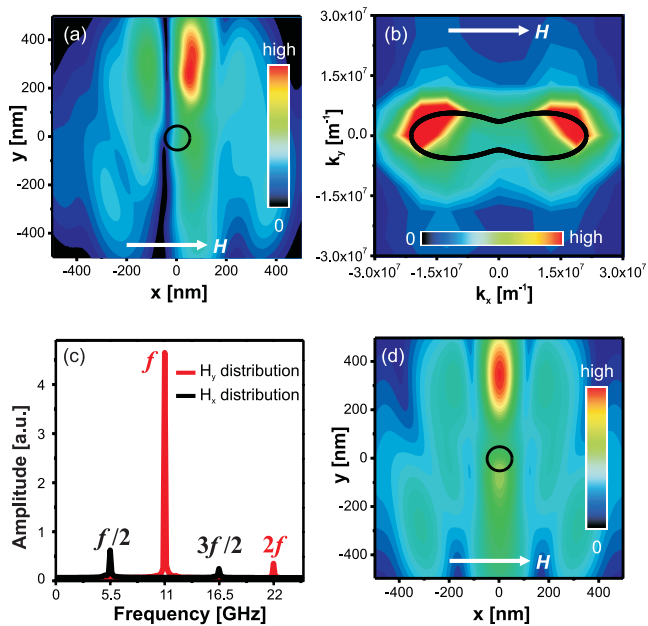}
\caption{\label{fig:halfmode} Spatial distribution of the
\emph{f}/2 mode obtained by taking into account in the
simulations: (a) both components ($H_{x}$, $H_{y}$) of the Oersted
field; (d) only the parallel component $H_{x}$. The black circle
indicates the position of the nanocontact. (b) The extracted
distribution of the \emph{f}/2 mode in the wavevector space
considering components ($H_{x}$, $H_{y}$) of the Oersted field;
(c) Magnetization oscillation spectrum for the cell at the top
edge of the contact considering i) only $H_{x}$ component (black
line) and ii) only $H_{y}$ component (red line) distributions of
the ac Oersted field.}
\end{figure}

The radiation pattern extracted from the simulation for the
\emph{f}/2 mode is presented in Fig.~\ref{fig:halfmode}(a). The
black circle indicates the position of the nanocontact. One can
observe that the \emph{f}/2 mode is localized outside the contact
area in the corresponding regions of travelling spin waves for the
directly excited \emph{f} mode. This observation may suggest that
we are dealing with a splitting of magnons of the \emph{f} mode.
The law of energy conservation for such processes is fulfilled
since any mode could be obtained as a combination (confluence or
splitting) of the other modes (see Fig.~\ref{fig:spectrum}(a)).
However, it has to be noted that the wave numbers extracted from
the magnetization intensity distributions in the phase space for
both \emph{f} and \emph{f}/2 (Fig.~\ref{fig:fmode}(c) and
Fig.~\ref{fig:halfmode}(b)) modes do not fulfill the momentum
conservation law. Moreover, from the constant-frequency curves
calculated analytically for 11 GHz and 5.5 GHz (black and yellow
curves from Fig.~\ref{fig:fmode}(c)) one sees that the
conservation of momentum during the splitting process is not
satisfied for any combination of wavevectors. In other words,
there is no pair of wavevectors at the frequency of \emph{f}/2 =
5.5 GHz that can combine into one corresponding to 11 GHz, or vice
versa. This strongly contradicts the assumption that three-magnon
scattering processes take place in the given configuration, ruling
out mechanism 1. The analysis in the phase space of the \emph{f}/2
mode (displayed in Fig.~\ref{fig:halfmode}(b)) shows a propagation
of spin waves with wavevectors under a maximum angle of
$\sim$15$^{\circ}$ with respect to the bias field ($x$-direction).
The asymmetry is related to additional contributions of the lead
to the internal field (similar for the asymmetry in the radiation
pattern from Fig.~\ref{fig:halfmode}(a)).

In order to discriminate between mechanism 2 and 3 for the
\emph{f}/2 mode generation (perpendicular and parallel parametric
pumping), we performed two additional simulations, in each one
taking into account only one component of the Oersted field,
$H_{y}$ and $H_{x}$, respectively. The applied power was kept at
the same value as previously.

By applying only the perpendicular component $H_{y}$, the
qualitative behavior for the \emph{f} mode remains virtually
unchanged in both space and phase domain. Practically no
difference was observed in comparison with the two distributions
presented in Fig.~\ref{fig:fmode}(a) and Fig.~\ref{fig:fmode}(c).
This confirms that the \emph{f} mode derives from the
perpendicular component of the ac Oersted field. However, from the
frequency spectra (red curve in Fig.~\ref{fig:halfmode}(c)) it was
observed that the \emph{f}/2 mode had vanished. Even if the
pumping power was doubled, no peak was seen at the corresponding
frequency for the \emph{f}/2 mode. Therefore, one can conclude
that perpendicular parametric pumping (mechanism 2) is not
responsible for the generation of the \emph{f}/2 mode. Also, it
evidences that the splitting of magnons does not occur even if
there is a strong excitation of the \emph{f} mode.

In the longitudinal parametric pumping process (mechanism 3), the
magnetization is excited by an ac field parallel to the steady
one. Within this process, spin waves having half of the excitation
frequency are amplified directly from the thermal bath. The
frequency spectra obtained from the simulation (black line in
Fig.~\ref{fig:halfmode}(c)) show two peaks corresponding to the
\emph{f}/2 and 3\emph{f}/2 modes and no intensity for the \emph{f}
mode. The spatial distribution of the \emph{f}/2 mode is presented
in Fig.~\ref{fig:halfmode}(d). Two maxima are observed and are
consistent with the spatial distribution of the excitation field
(only $H_{x}$ component). The strong asymmetry is derived from the
additional field created by the lead on top of the sample. One can
notice that the intensity distribution of the \emph{f}/2 mode
obtained in this case is very similar to the one resulting when
both components of the Oersted field were taken into account
(Fig.~\ref{fig:halfmode}(a)). Moreover, the analysis of the
\emph{f}/2 mode in the phase space shows exactly the same
distribution of the wave numbers as was found when both $H_{x}$
and $H_{y}$ components were considered
(Fig.~\ref{fig:halfmode}(b)). The slight shift observed by
comparing the two magnetization distributions from
Fig.~\ref{fig:halfmode}(a) and Fig.~\ref{fig:halfmode}(c) can be
interpreted as a distortion caused by the travelling waves of the
\emph{f} mode on the \emph{f}/2 mode. Altogether, this shows
clearly that the \emph{f}/2 mode is excited as an independent
eigenmode by parametric longitudinal pumping.

In conclusion, we performed a micromagnetic study of the linear
and nonlinear spin-wave modes excited by microwave currents in a
nanocontact structure. We have shown that the dominant spin-wave
mode is driven by the dynamical Oersted field component
perpendicular to the static external field. The nonlinear
\emph{f}/2 mode derives from the parametric longitudinal pumping
effect (due alternating Oersted field component parallel to the
bias field). Spectral positions of those modes are determined both
in the space and frequency domains. The \emph{f} mode emitted from
the points localized at edge of the contact is propagating nearly
perpendicular to the bias field. The \emph{f}/2 mode is localized
outside the contact area and is propagating under a small angle
with respect to the steady field. These results are important for
the realization of nano-sized spin-wave antennas in magnon based
spintronic devices and can be used for the design of spin-wave
coupled arrays of nano-oscillators.

This work was supported by the European Commission within the
EU-MRTN SPINSWITCH (MRTN-CT-2006- 035327). We also thank A.
Slavin, V. Tiberkevich and M. Kostylev for fruitful discussions.

\end{document}